\pdfoutput=1
\documentclass[EPJ]{webofc}
\usepackage[varg]{txfonts}   
\usepackage[colorlinks=true]{hyperref}
\woctitle{INPC13}
\begin{document}
\title{Continuum TDHF calculation of Isoscalar and Isovector Giant Monopole Resonances}
%
%

\author{P. D. Stevenson\inst{1}\fnsep\thanks{\email{p.stevenson@surrey.ac.uk}} \and C. I. Pardi\inst{1}\fnsep
}

\institute{Department of Physics, University of Surrey, Guildford, Surrey, GU2 7XH, United Kingdom}

\abstract{%
We motivate and summarise some recent results in the application of formally exact boundary conditions in nuclear time-dependent Hartree-Fock calculations, making use of Laplace transformations to calculate the values of the wave functions at the boundaries.  We have realised the method in the case of giant monopole resonances of spherically-symmetric nuclei, and present strength functions of $^{16}$O and $^{40}$Ca using a simplified version of the Skyrme force, showing that no artefacts from discretisation occur as contaminants}

\maketitle
\section{Introduction}
\label{intro}
Many nuclear phenomena fall under the general title of collective motion.  Examples of collective motion such as giant resonances, rotational states, fusion and deep inelastic collisions may all be simulated computationally via solutions of the time-dependent Schr\"odinger equation, rendered in its mean-field form of the time-dependent Hartree-Fock equation \cite{Sim12}.  

Practical calculations of nuclear time-dependent Hartree-Fock begin from one or more static nuclear states calculated with standard Hartree-Fock, arranged in a coordinate space grid and given initial velocities or internal boosts according to the desired motion.  In all cases, the physical boundary conditions of the single-particle wave functions are that they go to zero at infinite distance.  In practice, the static initialisations have a zero value for the wave functions at the edge of a computational box, typically around a factor 2-10 times the nuclear radius.  This box boundary condition has little adverse effect on the static ground states of nuclei, thanks to the exponentially decreasing wave function beyond the range of the nuclear potential.  In the case of time-dependent collective motion, however, a sufficient amount of wave function is emitted from the nucleus to result in the zeroing of wave functions at the boundaries to have a deleterious effect on some observables \cite{Rei06}.  The principal problem with zeroing the wave function at boundaries in time dependent problems is that outgoing waves are reflected back, in the same way that a wave on a string is reflected by a fixed boundary.

A variety of techniques exist to deal better with the boundary conditions in the time-dependent case.  Previous applications in nuclear physics include the application of masking functions \cite{Rei06} or the use of complex optical potentials to absorb outgoing flux \cite{Nak05}.  Either of these techniques can function adequately, though the size of the masking or absorbing region beyond the physical box can become large and in principle the methods do not absorb equally at all energies.

We present the application of formally exact boundary conditions for the particular case of giant monopole resonances in spherical nuclei.  Since one can specialise to spherical symmetry in this case, we are able to compare with ``exact'' calculations performed in a very large box, such that outgoing flux does not encounter the boundary during the simulation time.

\section{Giant Monopole Resonances}
Giant monopole resonances are $L=0$ collective excitations in which the nuclear density oscillates in a radial mode, with contributions from all single particle states.  The isoscalar (IS) mode involves the neutron and proton states oscillating in phase, and the isovector (IV) mode involves their vibration in anti-phase.  Note that the normal modes of oscillation need not be (and in general are not) exactly of IS or IV nature.  However, isospin, as an approximately good quantum number in nuclei, is a useful classification of nuclear excitation modes, and reaction mechanisms often begin an oscillation in just one isospin mode.  Giant Monopole resonances date experimentally to the late 1970s \cite{Har77,SpethReview}, and have been a topic of ongoing theoretical study \cite{Bwo79,Wu99,Alm05,And13}, not least for their link to nuclear matter properties \cite{Bla80,Cen10,Dut12}

In applying exact boundary conditions to the case of giant monopole resonances within the time-dependent Hartree Fock framework, we use a zero-range version of the Skyrme interaction \cite{StoRei}, as commonly used in exploratory work \cite{Wu99}.  We take the parameters $t_0=-1090.0$ MeV fm$^{3}$ and $t_3=17288.0$ MeV fm$^6$ \cite{Wu99} for all calculations presented.  This simplified Skyrme interaction cannot be expected to give quantitatively correct results, but it serves to test the method.

\section{Boundary conditions}

The method we use to compute the wave functions at the boundary of the spatial box is formally exact, though implemented with some approximations.  It takes the time-dependent Schr\"odinger equation for a wave function, defined as a function of temporal and spatial coordinates, Laplace-transforms the temporal coordinate, and considers the transformed Schr\"odinger equation in a region outside the complicated and nucleus-dependent nuclear potential.  A general solution can be obtained and evaluated at the spatial boundary point, given the known form of the Coulomb and centrifugal parts of the potential.  The inverse Laplace transform then gives the desired wave function. The numerical cost comes in the result that the expression for the wave function at the boundary is non-local in time, and consequently requires, at each time, an integration from the start of the simulation to the time at hand.  In this proceedings article, it is not possible to give a sufficient mathematical summary, but we have published a full derivation for the case of Coulomb-less neutron states \cite{Par13}, to which readers are referred, and we are preparing a follow-up for the proton states for publication.  A complete description can also be found in the PhD thesis of one of the coauthors of the present work \cite{Chrissthesis}.

\section{Strength Functions}
\begin{figure}[tbh]
\centering
\includegraphics[width=11cm,clip]{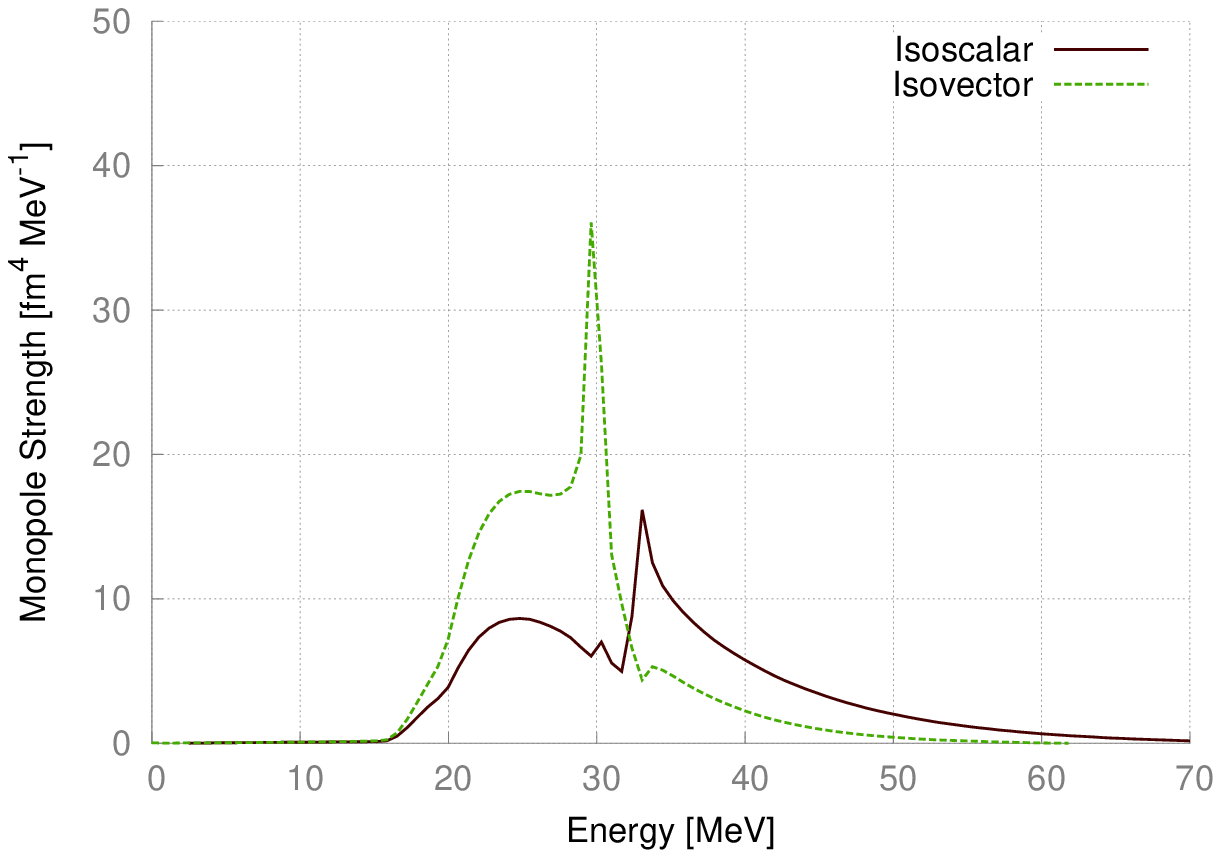}
\includegraphics[width=11cm,clip]{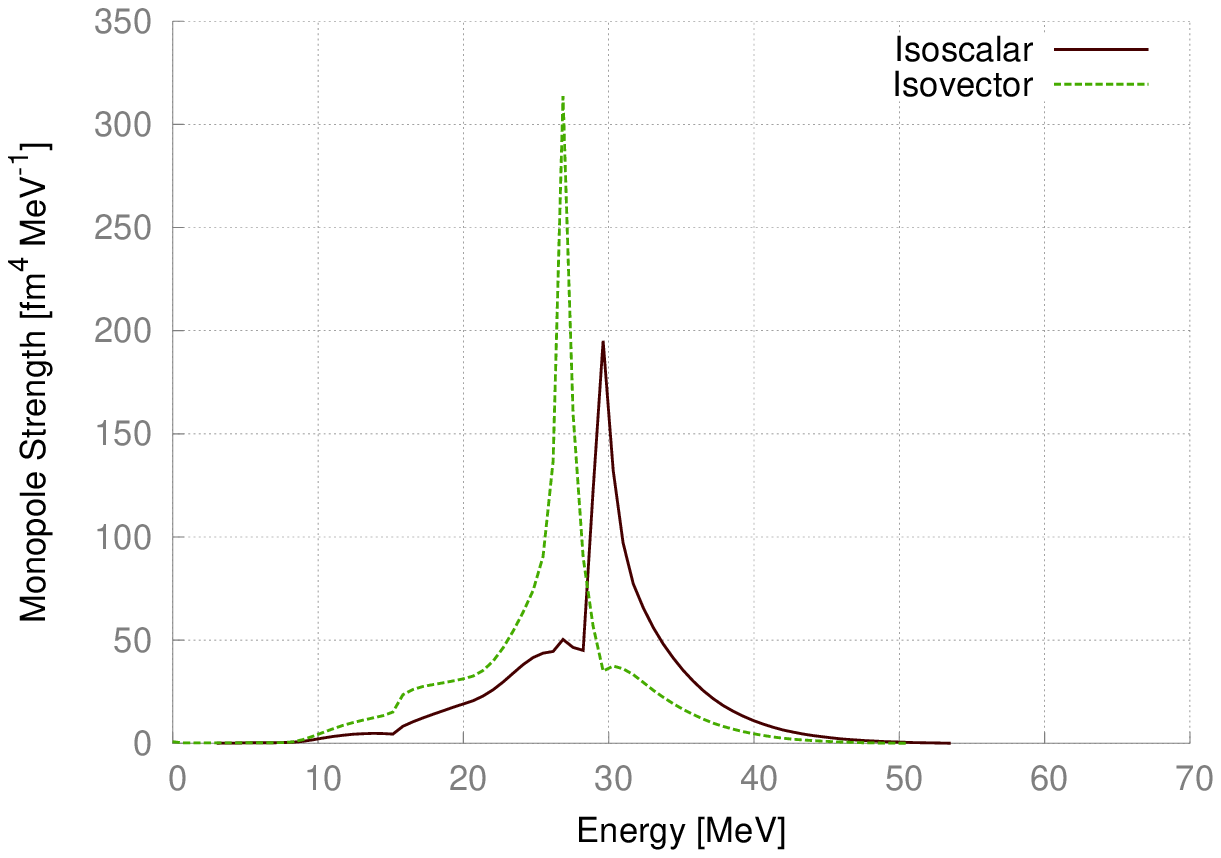}
\caption{Isovector and Isoscalar strength function in the case of $^{16}$O (upper panel) and $^{40}$Ca (lower panel) using a zero range version of the Skyrme interaction.  The energy resolution is 0.7 MeV.}
\label{fig-1}       
\end{figure}

The most physically relevant calculation to perform for giant resonances is the strength function, being related to a nucleus's probability of absorbing multipole strength at a given energy.  Usually defined as 
\begin{equation}
S(E) = \sum_\nu\left|\langle\nu|\hat{Q}|0\rangle\right|^2\delta(E-E_\nu),
\end{equation}
for an excitation mode given by operator $\hat{Q}$, it is calculated in time-dependent Hartree-Fock from the Fourier-transform of the time-dependent expectation value of $\hat{Q}(t)$ \cite{shimla}.  Once transformed, the quantum Ansatz $E=h\omega$ is used to interpret the time-dependent strength function in terms of energy.

As a selection of sample results, we present, in figure \ref{fig-1}, IS and IV giant monopole resonances for $^{16}$O and $^{40}$Ca for the present case of the simplified Skyrme force.  We note that in a small box, of radius 30fm) we are able to iterate much longer in time than a reflecting boundary will allow, and through the Nyquist theorem, therefore obtain a much more fine resolution in frequency, and hence energy.  The rough location of the peaks is consistent with other work \cite{And13,shimla}, though the relative position of the IS and IV peaks is reversed.  We repeat the comment that a good quantitative agreement with experiment, or more realistic theory, is not expected with this simplified force.

\section{Outlook and Conclusions}

We have presented time-dependent Hartree-Fock calculations with essentially analytic boundary conditions at large radius.  The method has allowed calculation of giant monopole resonances with a simplified Skyrme force.  In the short term, extension to the full Skyrme force is in progress, and unambiguously realisable.  In the longer term, the method can be applied to codes not requiring spherical symmetry - and so to deformed nuclear resonances, and collisions - though the computational time will be considerably longer.  We are working on such developments and in the longer term, analytic numerical boundary conditions have the scope to be applied to the most sophisticated nuclear time-dependent codes, covering all forms of collective motion.

%
%
%

\end{document}